\documentclass[twocolumn]{svjour3}         
\smartqed  
\usepackage{graphicx}
\begin{document}

\title{Cognitive computation with autonomously active neural networks:
an emerging field
}


\titlerunning{Autonomous Neural Networks}        

\author{Claudius Gros  
}


\institute{Claudius Gros \at
         Institute of Theoretical Physics\\
         J.W. Goethe University Frankfurt\\
         60054 Frankfurt/Main, Germany}

\date{Received: date / Accepted: date}

\maketitle

\begin{abstract}
The human brain is autonomously active. To understand 
the functional role of this self-sustained neural activity,
and its interplay with the sensory data input stream, is an
important question in cognitive system research and we 
review here the present state of theoretical modelling.

This review will start with a brief overview of the 
experimental efforts, together with a discussion of transient 
vs.\ self-sustained neural activity in the framework of
reservoir computing. The main emphasis will be then 
on two paradigmal neural network architectures showing 
continuously ongoing transient-state dynamics: 
saddle point networks and networks of attractor relics.

Self-active neural networks are confronted with two
seemingly contrasting demands: a stable internal
dynamical state and sensitivity to incoming stimuli.
We show, that this dilemma can be solved by networks
of attractor relics based on competitive
neural dynamics, where the attractor relics
compete on one side with each other for 
transient dominance, and on the other side 
with the dynamical influence of the input signals.

Unsupervised and local Hebbian-style online
learning then allows the system to build
up correlations between the internal dynamical
transient states and the sensory input stream.
An emergent cognitive capability results from this
set-up. The system performs online, and on its own,
a non-linear independent component analysis of 
the sensory data  stream, all the time being 
continuously and autonomously active. This 
process maps the independent components of 
the sensory input onto the attractor relics, 
which acquire in this way a semantic meaning.

\end{abstract}

\section{INTRODUCTION}

The brain has a highly developed and complex self-generated 
dynamical neural activity, and this fact raises a series
of interesting issues. Does this self-sustained neural
dynamics, its eigendynamics, have a central functional role, 
organizing overall cognitive computational activities? Or 
does this ongoing autonomous activity just serve as a kind 
of background with secondary computational task, 
like non-linear signal amplification or time encoding 
of neural codes?

The answer to this question is important not only
to system neurobiology, but also for research in
the field of cognitive computation in general. We will 
review here approaches based on the notion that the 
autonomous neural dynamics has a central regulating role
for cognitive information processing. We will 
then argue, that this line of research constitutes 
an emerging field in both computational neuroscience 
and cognitive system research.

Some preliminaries, before we start. This is a
mostly non-technical review with emphasis on content, 
an exhaustive and complete discussion of the published 
work on the subject is not the objective here. Centrally 
important equations will be given and explained, 
but for the numerical values of the parameters 
involved, and for the details of the simulation set-ups,
we will refer to the literature.
The discussion will be given generally from 
the perspective of cognitive system theory,
{\it viz} bearing in mind the overall
requirements of prospective complete cognitive systems,
akin to ones of real-world living animals \cite{grosBook2008,grosEmo09}.

\subsection{Autonomous brain dynamics}
\label{subsect_autonomous_brain_dynamics}

On the experimental side, the study of
self-induced or autonomous neural activity
in the brain has seen several developments
in recent years, especially by fMRI
studies \cite{fox2007}, and we will start 
by discussing some key issues arising in this 
respect.

The vast majority of experiments in cognitive
neuroscience study the evoked neural response to
certain artificial or natural sensory stimuli, 
often involving a given task which has been 
trained previously. It has been realized early 
on, that the neural response shows strong 
trial-to-trial variation, which is often as large 
as the response itself. This variability
in the response to identical stimuli is
a consequence of the ongoing internal
neural activities (for a discussion see
\cite{arieli1996}).
Experimentally one has
typically no control over the details of the
internal neural state and it is custom
to consider it as a source of noise,
averaging it out by performing 
identical experiments many times over.
It is on the other side well known that
the majority of energy consumption of the
brain is spent on internal processes \cite{raichle2006},
indicating that the ongoing and self-sustained
brain dynamics has an important functional role.
Two possibilities are currently discussed:

{\sl (A)} -- The internal neural activity could be 
in essence a random process with secondary functional
roles, such as non-linear signal amplification or
reservoir computing for the spatiotemporal encoding 
of neural signals (for a theory review see \cite{vogels2005}).

{\sl (B)} -- The internal neural activity could represent
the core of the cognitive information processing, being
modulated by sensory stimuli, but not directly and forcefully
driven by the input signals. Indications for this scenario
arise, e.g., from studies of the visual information 
processing \cite{fiser2004} and of the attention system \cite{fox2006}.

The overall brain dynamics is still poorly understood and
both possibilities (A) and (B) are likely to be relevant 
functionally in different areas. In this review we will
focus on the ramifications resulting from the second
hypothesis. There are indications, in this regard, that 
distinct classes of internal states generated autonomously
correspond to dynamical switching cortical states, 
and that the time series of the spontaneous
neural activity patterns is not random but determined
by the degree of mutual relations \cite{kenet2003}. Additionally,
these spontaneous cortical state may be semantic in nature, having a
close relation to states evoked by sensory stimuli \cite{ringach2003}
and to neural activity patterns induced via thalamic stimulation \cite{maclean2005}.
A second characteristics recurrently found in experimental
studies is the organization of the spontaneously active states
into spatially anticorrelated networks \cite{fox2005},
being transiently stable in time, in terms of firing rates,
with rapid switching between subsequent states \cite{abeles1995}.

These results indicate that certain aspects of the time evolution 
of the self-sustained neural activity in the brain have the form of 
transient state dynamics, which we will discuss in detail 
in Sect.\ \ref{sect_transient_state_dynamics}, together with a high
associative relation between subsequent states of mind. This
form of spontaneous cognitive process has been termed
`associative thought process' \cite{gros2005}. 

It is currently under debate which aspects of the
intrinsic brain dynamics is related to consciousness.
The global organization of neural activity in anticorrelated
and transiently stable states has been suggested, on one
side, to be of relevance also for the neural foundations of
consciousness \cite{edelman2000,edelman2003},
{\it viz} the `observing self' \cite{baars2003a}.
The persistent default-mode network (for a critical 
perspective see \cite{morcom2007}), {\it viz} the
network of brain areas active in the absence of
explicit stimuli processing and task performance, 
has been found, on the other side, to be active
also under anesthetization \cite{vincent07}
and light sedation \cite{greicius2008}.
It is interesting to note, in this context, that
certain aspects of the default resting mode
can be influenced by meditational practices \cite{pagnoni2008}.

\begin{figure*}[tb]
\centerline{\hfill
\includegraphics[width=0.75\textwidth]{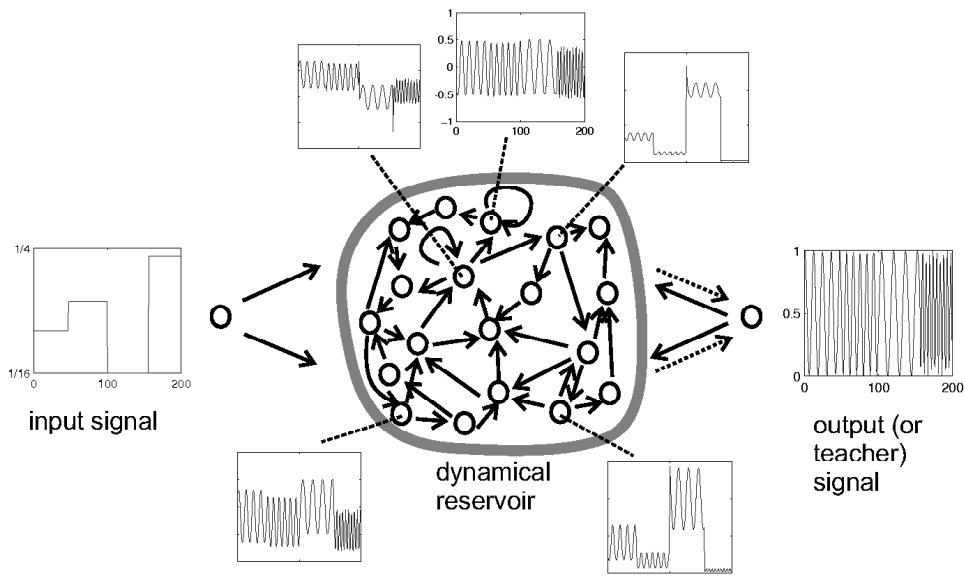}
           \hfill}
\caption{Operating principle of reservoir dynamics.
         The reservoir is activated by the input signal,
         mapping it in time and space to a higher-dimensional
         space. The reservoir activity provides then the basis for linear
         and supervised training of the output units, typically for
         time prediction tasks (figure courtesy H.~Jaeger).
        }
\label{figure_reservoir_dynamics}
\end{figure*}

\subsection{Reservoir computing}
\label{subsec_reservoir_computing}

The term `neural transients' characterizes
evoked periods of neural activities, remaining
 transiently stable after the disappearance 
of the primary stimulating signal. In the 
prolonged absence of stimuli, neural architectures
based on neural transients relax back to the 
quiescent default state. Network setups
based on neural transients therefore occupy a
role functionally in between pure 
stimulus-response architectures and systems 
exhibiting continuously ongoing
autonomous neural activity. An important
class of neural architectures based on neural
transients are neural reservoirs, which we
discuss now briefly.

A recurrent neural net is termed a reservoir,
if it is not involved in the primary cognitive 
information processing, having a supporting role.
A typical architecture is illustrated
in Fig.\ \ref{figure_reservoir_dynamics}. The
reservoir is a randomly connected network of 
artificial neurons which generally has only
a transiently stable activity in the absence of 
inputs, {\it viz} the reservoir has a 
short-term memory.

In the standard mode of operation an
input signal stimulates the network, giving raise
to complex spatiotemporal reservoir activities.
Normally, there is no internal learning inside
the reservoir, the intra-reservoir synaptic
strengths are considered fixed. Time prediction is
the standard application range for reservoir
computing. For this purpose the reservoir is
connected to an output layer and the activities
of the output neurons are compared to a
teaching signal. With supervised learning,
either online or off-line, the links leading 
from the reservoir to the output then acquire a
suitable synaptic plasticity.

There are two basic formulations of reservoir
computing. The `echo-state' approach using
discrete-time rate-encoding neurons \cite{jaeger2001,jaeger2004},
and the `liquid state machine' using
continuous-time spiking neurons \cite{maass2002,maass2004}.
In both cases the dimensionality of the input signal,
consisting normally of just a single line, is small 
relative to the size of the reservoir, which
may contain up to a few hundred neurons. Many 
nonlinear signal transformations are therefore 
performed by the reservoir in parallel 
and the subsequent perceptron-like 
output neurons may solve complex tasks via 
efficient linear learning rules.

Neural reservoirs are possible candidates for local
cortical networks like microcolumns. The bare-bones
reservoir network is not self-active, but feedback 
links from the output to the reservoir may stabilize
ongoing dynamical activity \cite{maass2007}. 
In any case, reservoir nets 
are examples of network architectures of type (A),
as defined in the previous section. The task of
the reservoir, non-linear signal transformation, is
performed automatically and has no semantic content.
All information is stored in the efferent synaptic links.

There is an interesting similarity, on a functional level,
of reservoir computing with the notion of a 
`global workspace' \cite{baars2003b,dehaene2003}. The global
workspace has been proposed as a global distributed computational 
cortical reservoir, interacting with a multitude of peripheral 
local networks involving tasks like sensory preprocessing or 
motor output. The global workspace has also been postulated to have 
a central mediating role for conscious processes \cite{baars2003b,dehaene2003},
representing the dominating hub nodes of a large-scale,
small-world cortical network \cite{shanahan2007}.

\begin{figure*}[tb]
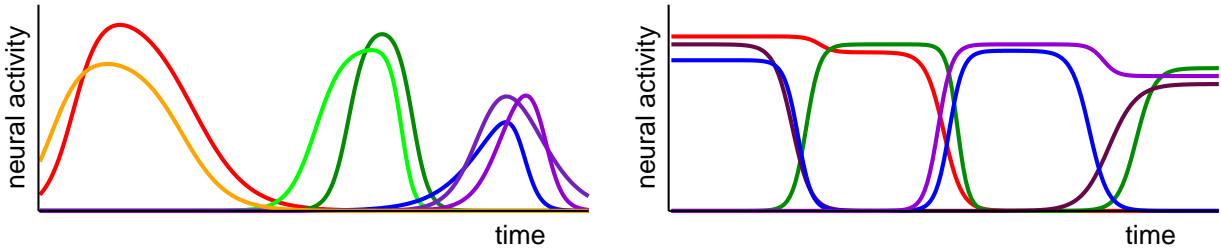

\centerline{\hfill
\includegraphics[width=0.45\textwidth]{transientStates_bump.eps}\hfill
\includegraphics[width=0.45\textwidth]{transientStates_slow.eps}
           \hfill}
\caption{Two examples of transient state dynamics. Left: Bump-like
phases of activities typically result from trajectories passing
close to saddle points. Right: Transient states with pronounced
plateaus are typical for multi-winners-take-all set-ups in the
context of networks with attractor relics.
        }
\label{figure_transient_states}
\end{figure*}

\section{TRANSIENT STATE DYNAMICS}
\label{sect_transient_state_dynamics}

A central question in neuroscience regards the neural code,
that is the way information is transmitted and encoded
(see \cite{shadlen1998,eggermont1998} for reviews). 
Keeping in mind that there is probably no pure information 
transmission in the brain, as this would be a waste of 
resources, that information is also processed when 
transmitted, one may then distinguish two issues regarding 
the encoding problem.

On one hand there is the question on how sensory signals
are reflected, on relative short timescales, in 
subsequent neural activities. Available neural degrees 
of freedom for this type of short-time encoding are the 
average firing rates (rate encoding), transient bursts 
of spikes and the temporal sequence of spikes 
(temporal encoding). In addition, the response of 
either individual neurons may be important, or the response
of local ensembles \cite{eggermont1998,averbeck2004}.

The subsequent sensory signal processing, 
on time\-scales typically exceeding 25-100ms,
may, on the other hand, involve neural dynamics
in terms of transiently stable activity patterns,
as discussed earlier in 
Sect.\ \ref{subsect_autonomous_brain_dynamics}.
In Fig.\ \ref{figure_transient_states} two
types of model transient state activities are
illustrated. Alternating subsets of neurons are 
either active, to various degrees, or essentially 
silent, resulting in well characterized transient
states having a certain degree of discreteness.
This discreteness should be reflected, on a higher 
level, on the properties of the corresponding
cognitive processes. Of interest in this context 
is therefore the ongoing discussion,
whether visual perception is continuous or 
discrete in the time domain \cite{vanrullen2003,kline2004},
on timescales of the order of about 100ms, with the
discrete component of perception possibly 
related to object recognition \cite{vanrullen2006}.
Transient state dynamics in the brain may therefore
be related to semantic recognition, a connection
also found in models for transient state
dynamics based on competitive neural dynamics.
In the following we will examine the occurrence
and the semantic content of autonomous transient
state dynamics in several proposed
cognitive architectures.

\begin{figure*}[tb]
\centerline{\hfill
\includegraphics[width=0.40\textwidth]{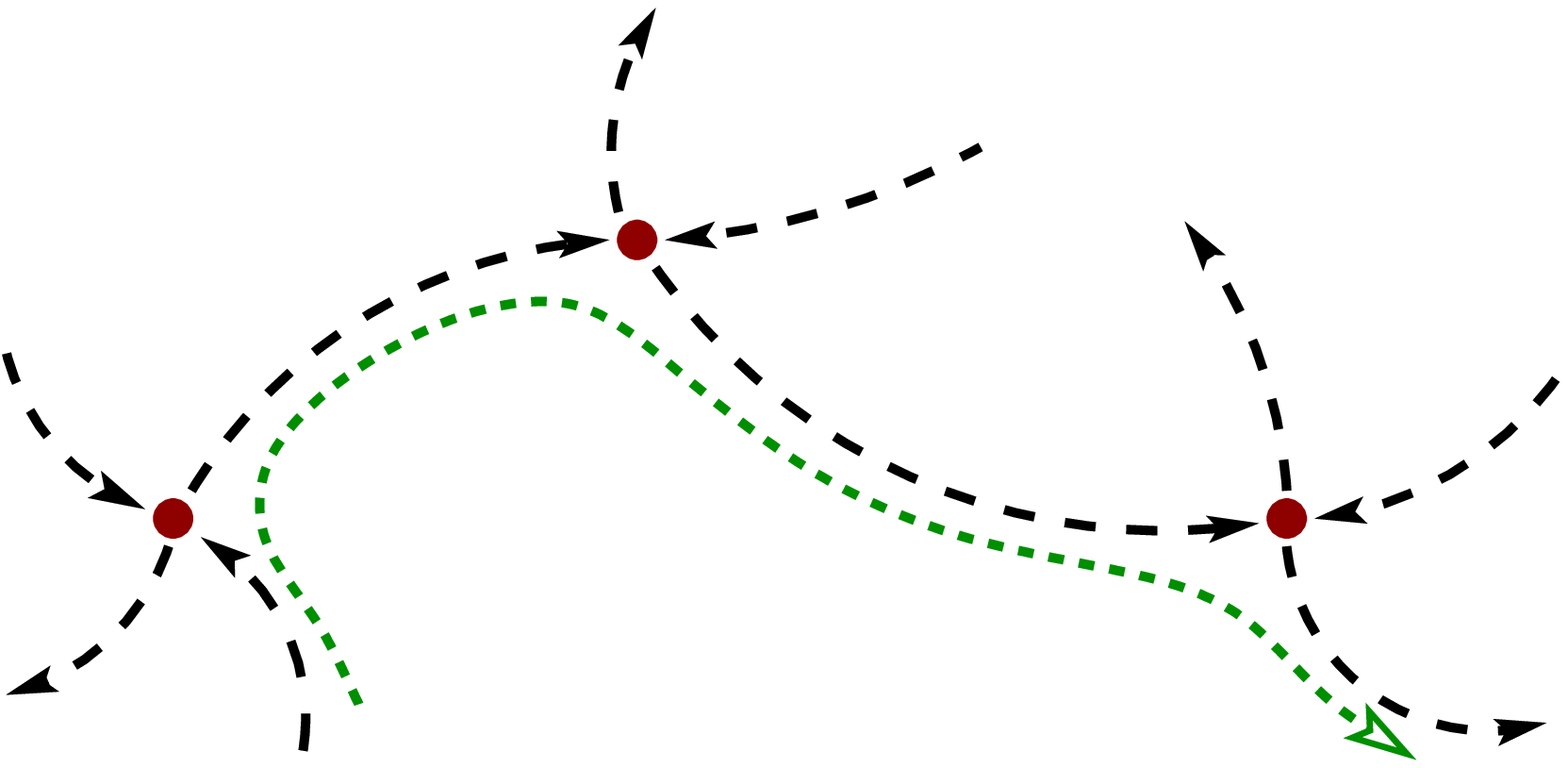}\hfill
\includegraphics[width=0.55\textwidth]{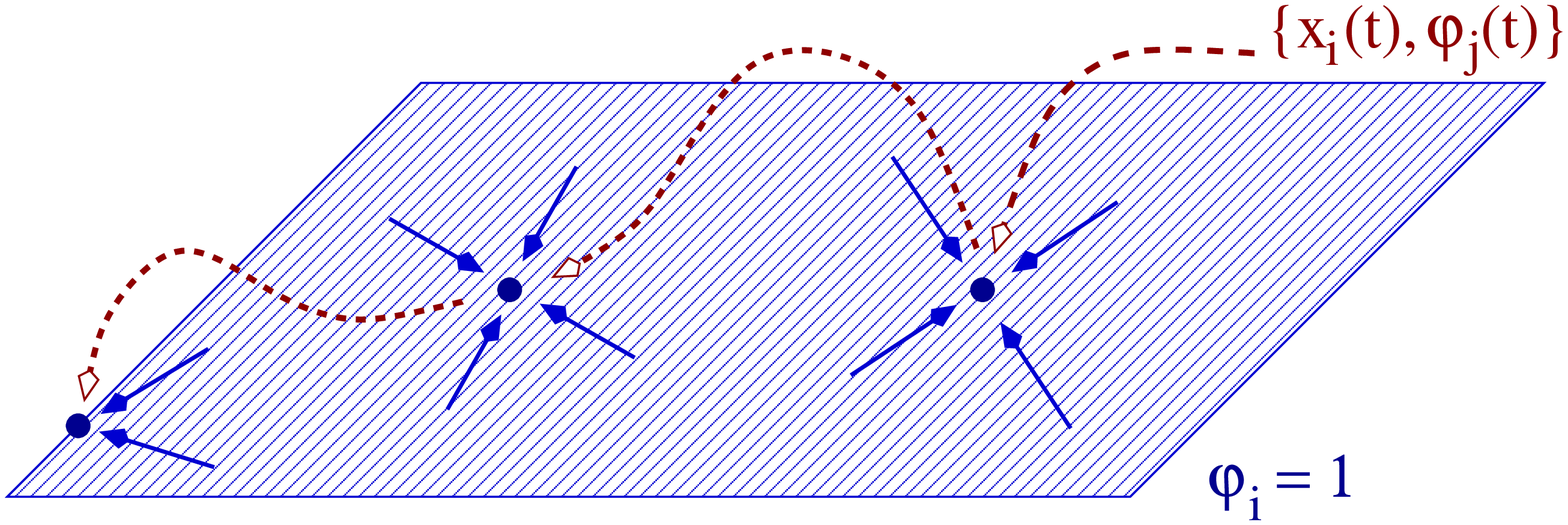}
           \hfill}
\caption{Two scenarios giving rise to transient state dynamics.
Left: A network of saddle points (filled circles) connected via 
heteroclines (long dashed lines) with a sample trajectory (short-dashed
line). The dynamics slows down close to a saddle point. 
Right: An attractor network (shaded plane) is embedded in a higher 
dimensional space via additional reservoir variables $\{\varphi_j(t)\}$,
leading to a destruction of the original fixpoints (filled circles), 
which are turned into attractor relics. The dynamics 
$\{x_i(t),\varphi_j(t)\}$ (short-dashed line) slows down
in the vicinity of an attractor relic.
        }
\label{figure_scenarios}
\end{figure*}

\subsection{Saddle point networks and winnerless competition}
\label{subsec_saddle_point_networks}

The concept of saddle point networks is based on the premises,
(a) that the internal ongoing autonomous dynamics organizes
the cognitive computation and (b) that the cognitive 
behavior is reproducible and deterministic in identical 
environments \cite{rabinovich2008}. As we will discuss in
the next section, the first assumption is shared with
attractor relic networks, while the second is not.

Technically, one considers a dynamical system, {\it viz} a
set of $N$ first-order differential equations and the set of 
the respective saddle points, compare Fig.\ \ref{figure_scenarios}.
The precondition is now that every saddle point has only
a single unstable direction and $(N-1)$ stable directions.
Any trajectory approaching the saddle point will then leave 
it with high probability close to the unique unstable separatrix
and the system therefore has a unique limiting cycle attractor.
This limiting cycle does not need to be a global attractor,
but normally has a large basin of attraction. During one passage
most, if not all, saddle points are visited one after the other,
giving raise to a transient state dynamics illustrated in
Fig.\ \ref{figure_transient_states}, with the trajectory slowing
down close to a saddle point.

Another condition for this concept to function is 
the formation of a heteroclinic cycle, which is of a set in
phase space invariant under time evolution \cite{krupa1997}.
Implying, as illustrated in Fig.\ \ref{figure_scenarios}, that the
unstable separatrix of a given saddle point needs to end up as
a stable separatrix of another saddle point. Such a behavior
occurs usually only when the underlying differential equations
are invariant under certain symmetry operations, like the exchange 
of variables \cite{krupa1997}. For any practical application, these
symmetries need to be broken and the limiting cycle will
vanish together with the heteroclinic sequence. It can however 
be restored in form of a heteroclinic channel, if the strength of the
symmetry-breaking is not too strong, by adding a stochastic
component to the dynamics. With noise, a trajectory loitering
around a saddle point can explore a finite region of phase 
space close to the saddle point until it finds the unstable
direction. Once the trajectory has found stochastically the
unstable direction, it will leave the saddle point quickly
along this direction in phase space and a heteroclinic 
channel is restored functionally. Cognitive 
computation on the backbone of saddle point networks 
is therefore essentially based on an appropriate 
noise level.

Cognitive computation with saddle point networks has been
termed `winnerless competition' in the context of
time encoding of natural stimuli \cite{rabinovich2001}
and applied to the decision making problem. In the later
case interaction with the environment may generate a second
unstable direction at the saddle points and decision taking
corresponds to the choice of unstable separatrix taken
by the trajectory \cite{rabinovich2008}.

\subsection{Attractor relic networks and slow variables}
\label{subsec_attractor_relic_networks}

A trivial form of self-sustained neural
activity occurs in attractor networks \cite{amit89}. Starting
with any given initial state the network state will
move to the next attractor and stay there, with
all neurons having a varying degree of constant
firing rates, the very reason attractor nets have 
been widely discussed as prototypes for the 
neural memory \cite{hasselmo99}.
As such, an attractor network is useless for 
a cognitive system, as it needs outside
help, or stimuli from other parts of the system, to
leave the current attractor. 

There is a general strategy which transforms an attractor
network into one exhibiting transient state dynamics,
with the transient neural states corresponding to the
fixpoints of the original attractor network. This
procedure is applicable to a wide range of attractor
networks and consists in expanding the phase space
by introducing additional local variables
akin to local activity reservoirs \cite{gros2005}.

To be concrete, let us denote with $\{x_i\}$ the
set of dynamical variables of the attractor network,
as illustrated in Fig.\ \ref{figure_scenarios}, and
by $\{\varphi_j\}$ the additional reservoir variables.
We assume that the reservoirs are depleted / filled
when the neuron is active / inactive,
\begin{equation}
T_\varphi\, \dot\varphi_j(t)\ \to\ \left\{
\begin{array}{rcl}
<0 &\quad& {\rm active\ neuron}\ j \\
>0 &\quad& {\rm inactive\ neuron}\ j 
\end{array}
                   \right.
~.
\label{eq_general_dot_phi}
\end{equation}
Together with a suitable coupling of the reservoir
variables $\{\varphi_j\}$ to the neural activities
$\{x_i\}$ one can easily achieve that the fixpoints
of the attractor networks become unstable, {\it viz}
that they are destroyed, turning into attractor
ruins or attractor relics. 

This situation is illustrated in Fig.\ \ref{figure_scenarios}. 
In the expanded phase space $\{x_i,\varphi_j\}$
there are no fixpoints left. It is not the case that
the attractors would just acquire additional unstable
directions, upon enlargement of the phase space,
turning them into saddle points. Instead, the enlargement
of the phase space destroys the original attractors
completely. The trajectories will however still slow down
considerably close to the attractor ruins, as illustrated
in Fig.\ \ref{figure_transient_states}, if the 
reservoirs are slow variables, changing only relatively
slowly with respect to the typical time constants
of the original attractor network. In this case the
time constant $T_\varphi$ entering the time evolution
of the reservoir, Eq.\ (\ref{eq_general_dot_phi}),
is large. In the limit $T_\varphi\to\infty$ the
reservoir becomes static and the dynamics is reduced
to the one of the original attractor network.

The dynamics exhibited by attractor relic networks is 
related to the notion of chaotic itinerancy \cite{tsuda2002},
which is characterized by trajectories wandering around
chaotically in phase space, having intermittent transient
periods of stability close to attractor ruins.
Here we consider the case of attractor relics arising
from destroyed point attractors. In the general case
one may also consider, e.g., limit cycles or 
strange attractors. 

The coupling to slow variables outlined here is a standard 
procedure for controlling dynamical systems \cite{grosBook2008},
and has been employed in various fashions for the
generation and stabilization of transient state dynamics.
One possibility is the use of dynamical thresholds for
discrete-time rate-encoding neural nets \cite{horn1989}.
In this case one considers as a slow variable the 
sliding-time averaged activity of a neuron and the
threshold of a neuron is increased / decreased 
whenever the neuron is active / inactive for a prolonged 
period. Another approach is to add slow components to
all synaptic weights for the generation of an externally 
provided temporal sequence of neural patterns \cite{sompolinsky1989}.
In the following we will outline in some detail an approach
for the generation of transient state dynamics which takes
an unbiased clique encoding neural net as its 
starting point \cite{gros2007}, with the clique encoding 
network being a dense and homogeneous associative network
(dHan).

\begin{figure*}[tb]
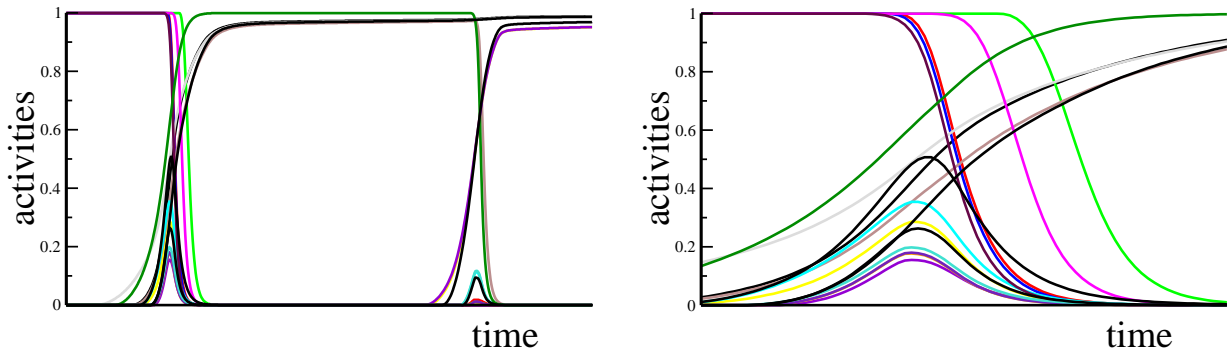

\centerline{\hfill
\includegraphics[width=0.45\textwidth]{18sitesCompetition.eps} \hfill
\includegraphics[width=0.45\textwidth]{18sitesCompCloseUp.eps}
           \hfill}
\caption{Multi-winners-take-all transient state dynamics and
         neural competition. The right graph is a 
         blow-up of the first transition in the left graph.
         During the transition many neurons compete with each other
         for becoming a member of the next winning coalition.
         The duration of the transient-state plateaus is
         given by the relaxation time $T_\varphi$ of the slow variables,
         see Eq.\ (\ref{eq_general_dot_phi}).
        }
\label{figure_competition}
\end{figure*}

\section{COMPETITIVE NEURAL DYNAMICS}

Transient state dynamics is intrinsically 
competitive in nature. When the current transient
attractor becomes unstable, the subsequent transient
state is selected via a competitive process.
Transient-state dynamics is a form of `multi-winners-take-all'
process, with the winning coalition of dynamical
variables suppressing all other competing activities
\cite{maass2000,oreilly1998}. Competitive proces\-ses
resulting in quasi-stationary states with intermittent
burst of changes are widespread, occurring in
many spheres of the natural or the social sciences. In 
the context of Darwinian evolution, to give an example, 
this type of dynamics has been termed 
`punctuated equilibrium' \cite{grosBook2008}. In the context
of research on the neural correlates of consciousness,
these transiently stable states in form of winning 
coalitions of competing neural ensembles have been proposed
as essential building blocks for human states of the
mind \cite{crick2003,koch2004}.

The competitive nature of transient state dynamics is
illustrated in Fig.\ \ref{figure_competition}, where
a representative result of a simulation for a dHan net
is presented. During the transition from one winning
coalition to the subsequent, many neurons try to become
members of the next winning coalition, which in the
end is determined by the network geometry, the synaptic
strengths and the current reservoir levels of the
participating neurons. 

The transition periods from one transient state to the 
next are periods of increased dynamical sensibility. When 
coupling the network to sensory inputs, the input signal
may tilt the balance in this competition for the next
winning coalition, modulating in this way the ongoing
internal dynamical activity. Transient state dynamics
therefore opens a natural pathway for implementing
neural architectures for which, as discussed in the introduction,
the eigendynamics is modulated, but not driven, 
by the sensory data input stream. A concrete example 
of how to implement this procedure will be discussed 
in Sect.\ \ref{sect_influence_stimuli}.

\begin{figure*}[tb]
\centerline{\hfill
\includegraphics[width=0.40\textwidth]{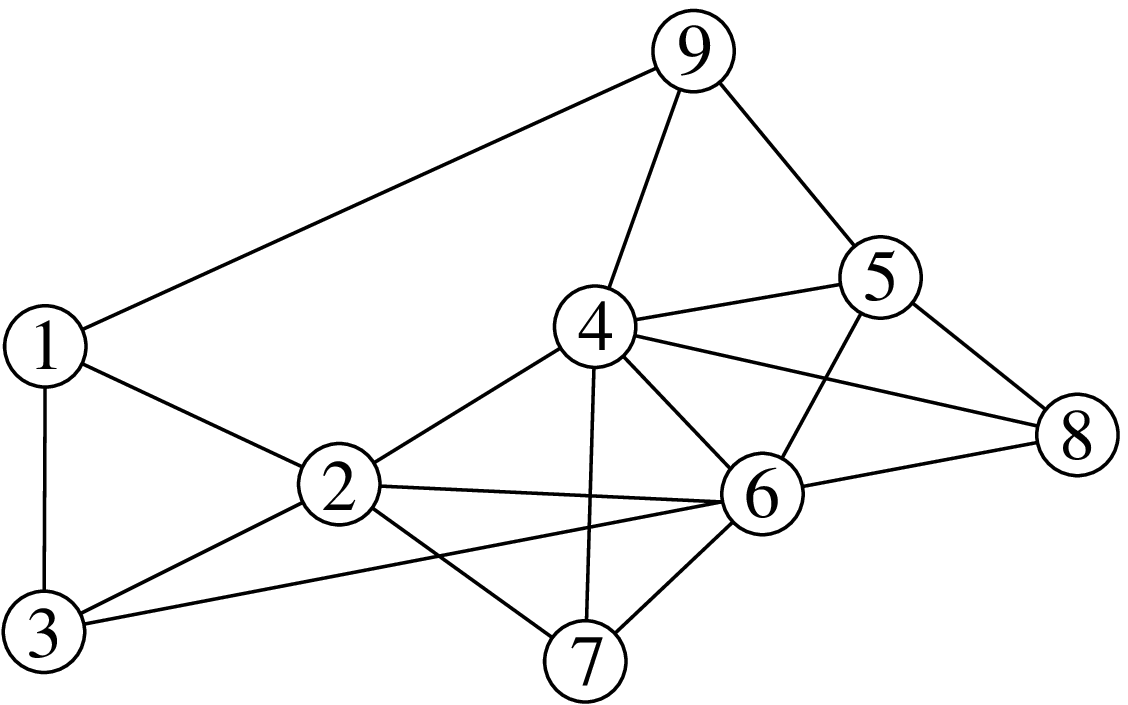} \hfill
\includegraphics[width=0.45\textwidth]{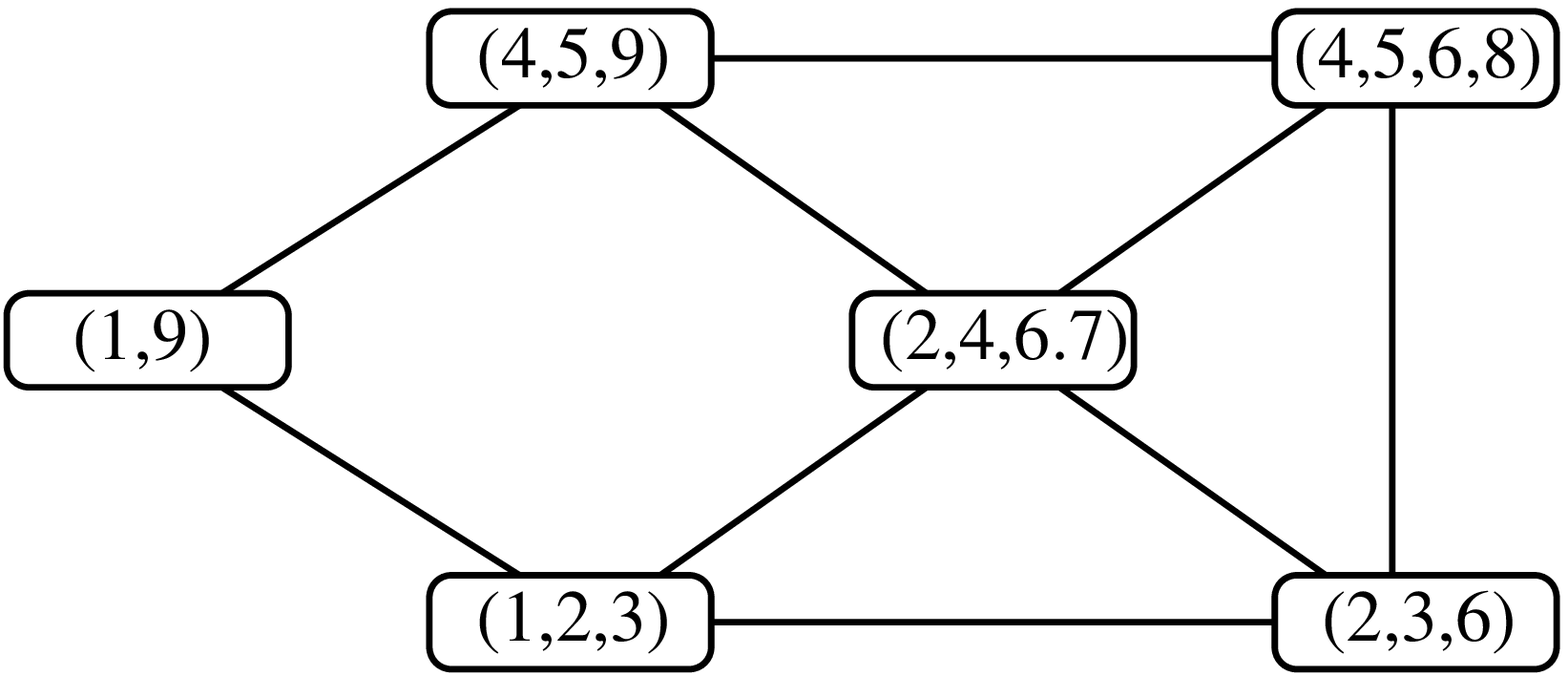}
           \hfill}
\caption{Clique encoding. On the left a 9-site network.
On the right the 5 cliques contained in the left-side network
are given. The cliques are connected via edges whenever they
share one or more sites and can therefore be considered as the
constituent vertices of a meta network of cliques.
        }
\label{figure_clique_encoding}
\end{figure*}

\subsection{Clique encoding}

Only a small fraction of all neurons are active at any
time in the brain in general, and in areas important for the
memory consolidation in particular \cite{quiroga2008}.
For various reasons, like the 
optimization of energy consumption and the maximization 
of computational capabilities \cite{olshausen2004}, 
sparse coding is an ubiquitous and powerful 
coding strategy \cite{maass2000}. Sparse coding may
be realized in two ways, either by small non-overlapping
neural ensembles, as in the single-winner-take-all architecture,
or by overlapping neural ensembles. The latter pathway draws 
support from both theory considerations \cite{quiroga2008},
and from experimental findings.

Experimentally, several studies of the Hippocampus indicate
that overlapping neural ensembles constitute
important building blocks for the real-time encoding 
of episodic experiences and representations \cite{lin2005,lin2006}. 
These overlapping representations are not random
superpositions but associatively connected. A 
hippocampal neuron could response, e.g., to various 
pictures of female faces, but these pictures would 
tend to be semantically connected, e.g.\ they
could be the pictures of actresses from the 
same TV series \cite{quiroga2005}. It is therefore
likely that the memory encoding overlapping
representations form an associative network,
a conjecture that is also consistent with studies of
free associations \cite{nelson2004,palla2005}.

There are various ways to implement overlapping
neural encoding with neural nets. Here we discuss 
the case of clique encoding. The term clique stems
from graph theory and denotes, just as a clique
of friends, a subgraph where (a)
every member of the clique is connected with all
other members of the clique and where (b) all other
vertices of the graph are not connected to each member
of the clique. In Fig.\ \ref{figure_clique_encoding}
a small graph is given together with all of its cliques.

Also shown in Fig.\ \ref{figure_clique_encoding} are
the associative interconnections between the cliques. One
may view the resulting graph, with the cliques as
vertices and with the inter-clique associative
connections as edges, as a higher-level representation
of an implicit hierarchical object definition
\cite{riesenhuber1999}.
The clique (4,5,9) in the original graph in 
Fig.\ \ref{figure_clique_encoding} corresponds to
a primary object and the meta-clique
[(4,5,9)-(2,4,6,7)-(4,5,6,8)] in the graph of
the cliques would in this interpretation encode
a meta object, composed of the primary objects
(4,5,9), (2,4,6,7) and (4,5,6,8).
This intrinsic possibility of hierarchical object 
definitions when using clique encoding has however 
not yet be explored in simulations and may be of
interest for future studies.

Cliques can be highly overlapping and there can be
a very large number of cliques in any given graph
\cite{kaczor2008,grosKaczor2008}. We will construct now a neural net
where the cliques of the network are the attractors. 
It is a homogeneously random and dense 
associative network (dHan), where the associative 
relations between cliques are given by the number of
common vertices. Starting from this attractor 
network we will introduce slow variables, 
as discussed in Sect.\ \ref{subsec_attractor_relic_networks},
in terms of local reservoirs. The network will
then show spontaneously generated transient state
dynamics, with the neural cliques as the attractor ruins.
In a second step we will couple the dHan net to
sensory stimuli and study the interplay between
the internal autonomous dynamical activity and the
data input stream. We will find that the cliques acquire 
semantic content in this way, being mapped autonomously
to the statistically independent patterns of the data input stream.

\subsection{Attractor network}
\label{subsec_attractor_network}

The starting point of our considerations is the
underlying attractor network, for which we employ
a continuous time formulation, with rate encoding neurons, 
characterized by normalized activity levels
$x_i\in[0,1]$. The growth rates $r_i$
govern the respective time developments,
\begin{equation}
\dot x_i \ =\ 
\left\{
\begin{array}{rcl}
(1-x_i)\,r_i &\quad& (r_i>0) \\
x_i\, r_i &\quad& (r_i<0) 
\end{array}
\right. ~.
\label{eq_x_dot}
\end{equation}
When $r_i>0$, the respective neural activity $x_i$ increases, 
approaching rapidly the upper bound; when
$r_i<0$, it decays to zero. We split the
rates into three contributions:
\begin{equation} 
r_i \ = \ r_i^{(+)}\,+\,r_i^{(-)}\,+\, r_i^{(ext)}
~.
\label{eq_r_pos_neg_contribution}
\end{equation}
An internal positive contribution $r_i^{(+)}\ge0$,
an internal negative contribution $r_i^{(-)}\le0$,
and the influence of external stimuli,
$r_i^{(ext)}$. We will discuss the influence
of non-trivial external stimuli in
Sect.\ \ref{sect_influence_stimuli}, for the
moment we consider $r_i^{(ext)}\equiv0$.
The division into an exciting and a depressing 
contribution in Eq.\ (\ref{eq_r_pos_neg_contribution})
reflects on one side the well known
asymmetry between excitatory neurons and 
inhibitory interneurons in the brain \cite{arbib2002} 
and is on the other side essential for clique encoding.
The $r_i^{(+)}$ are determined via
\begin{equation}
r_i^{(+)} \ =\  f_w(\varphi_i) \sum_{j} w_{ij}\, x_j
\label{eq_r_i_plus} 
\end{equation}
by the influence of the excitatory synaptic weights,
$w_{ij}\ge0$. The function $f_w(\varphi)$
entering Eq.\ (\ref{eq_r_i_plus}) couples
the dynamics of the neurons locally to the slow
variables $\varphi_i$. We will examine the
reservoir function $f_w(\varphi)$ in the 
next Section. For the 
time being we set $f_w(\varphi)\equiv1$, the primary 
neural dynamics is then decoupled from the
reservoir dynamics and we will retain the
starting attractor network.
The $r_i^{(-)}\le 0$ are given by
\begin{equation}
r_i^{(-)} \ =\ z\,
\tanh\left( \sum_{j} z_{ij}\,x_j\,f_z(\varphi_j)
          \right) 
~,
\label{eq_r_i_neg} 
\end{equation}
where the $z_{ij}\le0$ are the inhibitory 
synaptic weights and where $z>0$ sets
the scale for the inhibition.
Eq.\ (\ref{eq_r_i_neg}) leads
to a normalization $-z< r_i^{(-)}\le 0$.
We postpone the discussion of the reservoir function
$f_z(\varphi)$ and consider for the
time being $f_z(\varphi)\equiv1$.

Clique encoding corresponds to a multi-winners-take-all 
formulation. An inhibitory background is therefore
necessary. The dHan architecture contains hence
an inhibitory link $z_{ij}$ whenever
there is no excitatory link $w_{ij}$,
\begin{equation}
z_{ij} \ =\ \left\{
\begin{array}{rcl}
-z &\quad& (w_{ij}=0) \\
0 &\quad&  (w_{ij}>0)
\end{array}
                   \right.
~,
\label{eq_z_shunting}
\end{equation}
{\it viz} the excitatory links are
shunting the inhibitory syn\-aps\-es.
This inhibitory background is implicitly
present for the 9-site network shown
in Fig.\ \ref{figure_clique_encoding}.
The edges of the network shown 
in Fig.\ \ref{figure_clique_encoding}.
correspond to excitatory links $w_{ij}>0$. All
pairs of sites not connected by an edge
in Fig.\ \ref{figure_clique_encoding}
inhibit each other via $z_{ij}<0$.

The formulation of the attractor network with
clique encoding is such complete \cite{gros2007}.
All members of a given clique excite each other via
intra-clique $w_{ij}>0$. Neurons which are
not members of the current active clique
are suppressed by at least one inhibitory
link $z_{ij}<0$. This suppression $\sim (-z)$,
compare Eq.\ (\ref{eq_r_i_neg}), dominates
the residual positive signal the out-of-clique
neuron may receive, whenever $z$ is large enough.

An interesting feature of the dHan architecture
is the absence of a bias in Eq.\ (\ref{eq_x_dot}). 
There is no self excitation
or suppression, $w_{ii}=z_{ii}=0$. The dynamics
of an individual neuron is exclusively driven by
the influence of the network, it has no preferred firing
state. This feature would correspond for real-world neurons 
to the existence of a background of afferent activities with
a level close to the firing threshold.

Next we note, that the separation of scales
$z\gg w_{ij}$ implies that Heb\-bian-type modification
of the inhibitory links $z_{ij}$ would be meaningless, small
changes of a relatively large quantity will not
lead to a substantial effect. Hebbian learning in
the dHan architecture is therefore operational only for
the excitatory links $w_{ij}$, in accordance to the
general assumption that most learning taking place
in the brain involves synapses interconnecting neurons
and not interneurons \cite{arbib2002}. In 
Sect.\ \ref{sect_influence_stimuli} we will
consider the synaptic plasticity of links afferent
to the dHan layer. Unsupervised and local Hebbian-style
learning can however be implemented easily
for the intra-dHan excitatory links $w_{ij}$
for unsupervised and homeostatic calibration of
the excitatory links \cite{gros2007}. It is however
not essential for the occurrence and for the
stabilization of transient state dynamics, our
focus here.

\subsection{Reservoir activity}
\label{subsec_reservoir activity}

We consider normalized slow variables
$\varphi_i\in[0,1]$, with the time evolution
\begin{equation}
\dot\varphi_i(t)\,=\, 
r_{\varphi_i}^{av}+ {1\over T_\varphi}
\left\{
\begin{array}{ccl}
-\varphi_i &\ &     {\rm active\ neuron}\ i \\
(1-\varphi_i) &\ &  {\rm inactive\ neuron}\ i 
\end{array}
                   \right.
~,
\label{eq_dot_phi}
\end{equation}
where a neuron is active / inactive whenever
its activity level $x_i$ is close to unity /
zero. The $\varphi_i$ behave functionally 
as reservoirs, being depleted / refilled for
active / inactive neurons. The term
$r_{\varphi_i}^{av}$ on the RHS of
Eq.\ (\ref{eq_dot_phi}) is not essential
for the establishment of transient state 
dynamics, but opens an interesting alternative
interpretation for the slow variables. 
$r_{\varphi_i}^{av}$ vanishes for inactive 
neurons and takes the value
\begin{equation}
r_{\varphi_i}^{av}\big|_{{\rm active}\ i} \ =\
{1\over T_\varphi^{av}}
\sum_{{\rm active}\ j} \left(\varphi_j-\varphi_i\right)
\label{eq_r_phi_av}
\end{equation}
for active neurons. The reservoir levels $\{\varphi_i\}$ of
all active neurons are drawn together consequently.
All members of the currently active winning coalition
have then similar reservoir levels after a short time,
on the order of $T_\varphi^{av}$. 
This is a behavior similar to
what one would expect for groups of spiking neurons
forming winning coalitions via synchronization of
their spiking times. For each neuron of the winning
coalitions one could define a degree of synchronization,
given by the extent this neuron contributes to the
overall synchronization. Initially, this degree of
synchronization would have a different value for
each participating neuron. On
a certain timescale, denoted here by $T_\varphi^{av}$,
the spiking times would then get drawn together,
synchronized, and all members of the winning coalition 
of active neurons would then participate to a similar 
degree in the synchronized firing. The firing of the 
winning coalition would however not remain coherent 
forever. Internal noise and external influences
would lead to a desynchronization on a somewhat longer
time scale $T_\varphi\gg T_\varphi^{av}$. When desynchronized, 
the winning coalition would loose stability, giving way to a new
winning coalition. In this interpretation the reservoirs
allow for a ``poor man's'' implementation of 
self organized dynamical synchronization of neural
ensembles, a prerequisite for the temporal coding 
hypothesis of neural object definition
\cite{vonDerMalsburg1999,singer1995}.

Finally we need to specify the reservoir coupling functions
$f_w(\varphi)$ and $f_z(\varphi)$ entering 
Eqs.\  (\ref{eq_r_i_plus}) and (\ref{eq_r_i_neg}).
They have sigmoidal form with
\begin{equation}
f_w(\varphi),\ f_z(\varphi)\ \to\ \left\{
\begin{array}{rcl}
\sim 1 &\quad& \varphi\to 1 \\
\sim 0 &\quad& \varphi\to 0
\end{array}
                   \right.
~,
\label{eq_res_functions}
\end{equation}
and a straightforward interpretation:
It is harder to excite a neuron with
depleted reservoir, compare Eq.\  (\ref{eq_r_i_plus}),
and a neuron with a low reservoir level
has less power to suppress other neurons,
see Eq.\  (\ref{eq_r_i_neg}). Reservoir functions
obeying the relation (\ref{eq_res_functions}) therefore
lead in a quite natural way to transient state dynamics.
On a short time scale the system relaxes towards the
next attractor ruin in the form of a neural clique. Their 
reservoirs then slowly decrease and when depleted they
can neither continue to mutually excite each other, nor
can they suppress the activity of out-of-clique neurons
anymore. At this point, the winning coalition becomes 
unstable  and a new winning coalition is selected via 
a competitive process, as illustrated in 
Fig.\ \ref{figure_competition}.

Any finite $T_\varphi<\infty$ leads to the destruction
of the fixpoints of the original attractor network,
which is thus turned into an attractor relic network.
The sequence of winning coalitions, given by the
cliques of the network, is however not random. Subsequent
active cliques are associatively connected. The
clique (1,9) of the 9-site network shown in 
Fig.\ \ref{figure_clique_encoding}, to give an example,
could be followed by either (4,5,9) or by
(1,2,3), since they share common sites. The competition
between these two cliques will be decided 
by the strengths of the excitatory links and by the 
history of previous winning coalitions. If one of the
two cliques had been activated recently, the constituent
sites will still have a depressed reservoir and resist
a renewed reactivation.

\begin{figure}[tb]
\centerline{\hfill
\includegraphics[width=0.48\textwidth]{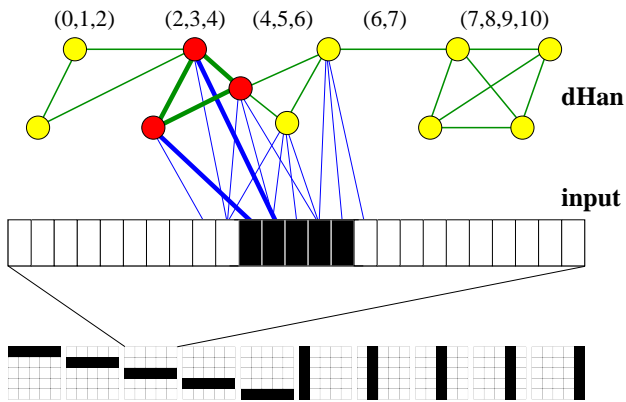}
           \hfill}
\caption{A dHan layer (top) with neural activities $\{x_i\}$
         and cliques (0,1,2), ... receives sensory signals
         via the input layer (middle) in the form of certain
         input patterns (bottom).
        }
\label{figure_dhan_input}
\end{figure}

The finite state dynamics of the dHan architecture is
robust. For the isolated network, we will discuss the
coupling to sensory input in the next section, the
dynamics is relaxational and dissipative \cite{gros2007}.
The system relaxes to the next attractor relic and
the reservoirs are relaxing either to zero or to unity,
depending on the respective neural activity levels. For 
a network with a finite number of sites, the long-time 
state will be a long limiting cycle of transient states.

The simulation results shown in Fig.\ \ref{figure_competition}
are for a set of parameters resulting in quite narrow
transitions and long plateaus \cite{gros2007}. The formulation
presented here allows for the modelling of the shape of the
plateaus and of other characteristics of the 
transient state dynamics. A smaller
$T_\varphi$ would result in shorter plateaus, a longer
$T_\varphi^{av}$ in longer transition times. One can,
in addition, adjust the shape of the reservoir functions
and details of  Eqs.\ (\ref{eq_r_i_plus}) and (\ref{eq_r_i_neg})
in order to tune the overall competition for the next winning coalition.
The dHan architecture providing therefore a robust 
framework for the generation of transient state dynamics,
offering at the same time ample flexibility and room
for fine tuning, paving the way for a range 
of different applications.

\section{INFLUENCE OF EXTERNAL STIMULI}
\label{sect_influence_stimuli}

The transient state dynamics generated by the
dHan architecture is dynamically robust. The
dHan dynamics has at the same time
windows of increased sensibility
to outside influences during the transition periods
from one transient state to the subsequent,
as shown in Fig.\ \ref{figure_competition}. These
transition periods are phases of active inter-neural 
competition, reacting sensibly to the influence
of afferent signals. 

We couple the input signals via an appropriate input
layer, as illustrated in Fig.\ \ref{figure_dhan_input},
denoting by $y_i\in[0,1]$ the time dependent input signals,
which we will take as black-and-white or grey-scaled
patterns. We denote by $v_{ij}$ the afferent links
to the dHan layer, with the external contribution to
the dHan-layer growth rates, compare
Eq.\ (\ref{eq_r_pos_neg_contribution}),
given by
\begin{eqnarray}
\label{eq_r_i_ext}
r_i^{(ext)} & =& \left\{
\begin{array}{ccl}
0                   &\quad & {\rm neuron}\ i\ {\rm active\ and} \ \Delta r_i<0 \\
\Delta r_i &\quad & 
\end{array}
                   \right. ~,
\\
\Delta r_i & =& \sum_j v_{ij}\, y_j ~.
\label{eq_delta_r_i}
\end{eqnarray}
The rationale behind this formulation is the following.
The role of the input signal is not to destabilize the
current winning coalition, the afferent signal is therefore
shunted off in this case, Eq.\ (\ref{eq_r_i_ext}). The 
input signal should influence the competition for the
next winning coalition, modulating but not driving
directly the dHan dynamics. This rational is realized by
the above formulation. Inactive neurons will receive
a bias $\Delta r_i$ from the input layer which 
increases / decreases its chance of joining the
next winning coalition for $\Delta r_i>0$
/ $\Delta r_i<0$.

\subsection{Novelty signal}
\label{subsec_novelity_signal}

A cognitive system with a non-trivial and 
self-sustained internal neural activity has to 
decide how and when correlations with the sensory 
data input stream are generated via correlations encoded
in the respective synaptic plasticities. This
is clearly a central issue, since the input data 
stream constitutes the only source for semantic 
content for a cognitive system. 

It makes clearly no sense if the
afferent links to the dHan layer,
{\it viz} the links leading from the
input to the internal network supporting
a self-sustained dynamical activity, would
be modified continuously via Hebbian-type
rules, since the two processes, the internal and
the environmental dynamics, are per se unrelated.
It makes however sense to build up correlation
whenever the input has an influence on the
internal activity, modulating the ongoing
associative thought process. From the perspective
of the cognitive system such a modulation of
the internal dynamics by environmental stimuli
corresponds to something novel and unexpected
happening. Novelty detection is therefore vital
for neural networks with a non-trival eigendynamics
processing sensory data. The importance of novelty detection 
for human cognition has been acknowledged indeed
 since long \cite{berns1997,barcelo2002}, and a possible
role of Dopamine, traditionally associated
with reinforcement reward transmission \cite{wise2004}, 
for the signalling of novelty 
has been suggested recently \cite{redgrave2006}.

\begin{figure}[tb]
\centerline{\hfill
\includegraphics[width=0.48\textwidth]{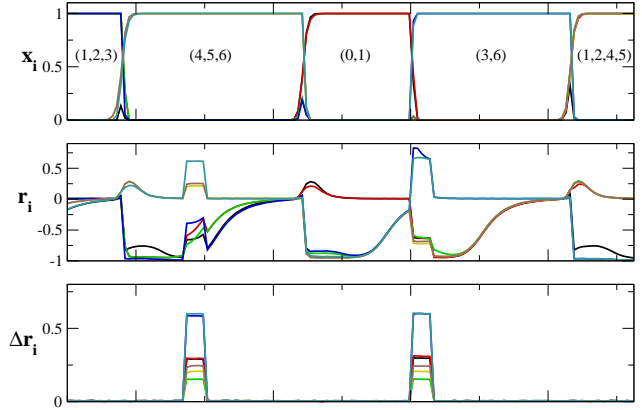}
           \hfill}
\caption{The activity levels $x_i$ of a dHan layer 
         containing seven neurons,
         compare Fig.\ \ref{figure_dhan_input},
         the growth rates $r_i$ and the contributions
         from the input-layer $\Delta r_i$, see
         Eq.\ (\ref{eq_delta_r_i}). The first input
         stimulus does not lead to a deviation of
         the transient state dynamics of the dHan layer.
         The second stimulus modulates the ongoing
         transient state dynamics, influencing the
         neural competition during the sensitive 
         phase. 
        }
\label{figure_input_yes_no}
\end{figure}

The influence of modulating and of not modulating sensory 
signals is illustrated in Fig.\ \ref{figure_input_yes_no},
where simulation results for a dHan layer containing
seven neurons coupled to an intermittent input signal
are presented. The signal is not able to
deactivate a currently stable winning coalition,
compare Eq.\ (\ref{eq_delta_r_i}), but makes an
impact when active during a transition period.
The system has the possibility to figure out
whenever the later has happened. When the
input signal $r_i^{(ext)}$ is relevant then
\begin{equation}
\big(r_i>0\big)\qquad \mbox{and}\qquad
\big(r_i^{(+)} + r_i^{(-)} <0\big)~.
\label{eq_condition_signal}
\end{equation}
In this case the internal contribution $r_i^{(+)} + r_i^{(-)}$
to the growth rate is negative and the input 
makes a qualitative difference. We may
therefore define a global novelty 
signal $S=S(t)$ obeying
%
\[
\dot S \ =\  \left\{
\begin{array}{rcl}
\phantom{-}1/T_S^+ &\qquad& (r_i>0) 
             \mbox{\ and\ } (r_i<r_i^{(ext)})\\
-1/T_S^- &\qquad& \mbox{otherwise} 
\end{array}
\right. ,
\]
%
where we have used Eq.\ (\ref{eq_r_pos_neg_contribution}),
$r_i^{(+)} + r_i^{(-)}=r_i-r_i^{(ext)}$, and
where a $\sum_i$ is implicit on the RHS of 
the equation. The novelty signal needs to be 
activated quickly, with $T_S^+ \gg T_S^-$. 
Learning then takes place whenever the novelty 
signal $S$ exceeds a certain threshold.

\begin{figure}[tb]
\centerline{
\includegraphics[width=0.30\textwidth]{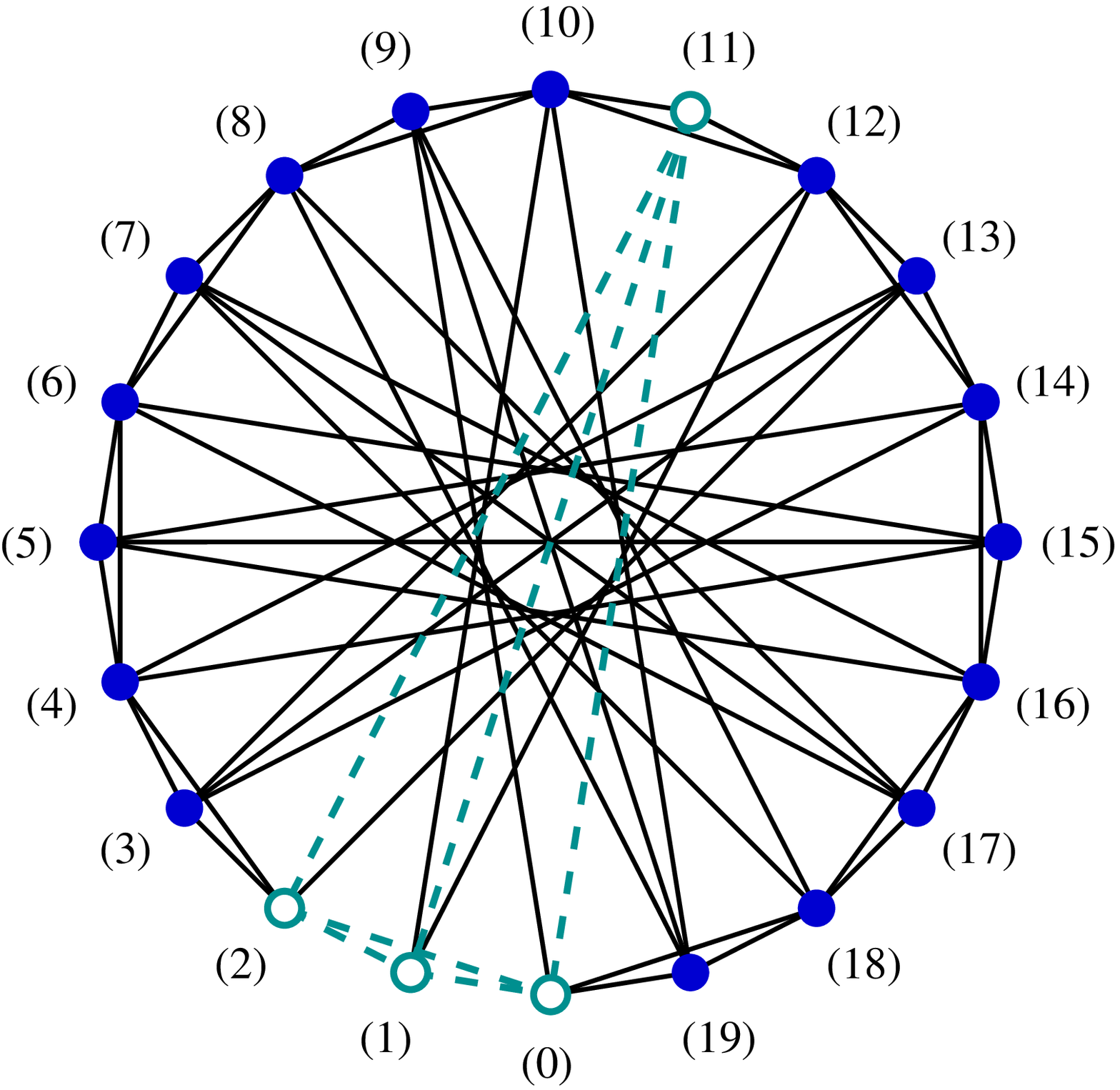}
           }
\vspace{1ex}
\centerline{
\includegraphics[width=0.48\textwidth]{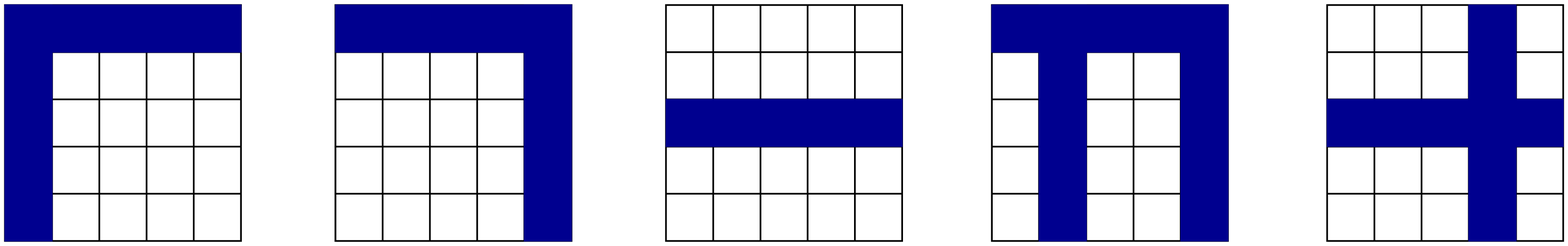}
           }
\caption{Top: The 20-site regular dHan net used for the bars
problem. It contains 10 cliques with four vertices. The
clique (0,1,2,11) is highlighted. Bottom: Some illustrative
input patterns for the $5\times5$ bars problem.
        }
\label{figure_star_pattern}
\end{figure}
\subsection{Afferent link plasticity: optimization principle}
\label{subsec_afferent_link_plasticity}

Having determined when learning takes place,
we have now to formulate the rules governing
how learning modifies the links afferent to the
dHan layer. For this purpose we will 
use the Hebbian principle, that positive
interneural correlations are enforced and
negative correlations weakened. Our system
is however continuously active, at no point
are activities or synaptic strengths reset.
The Hebbian principle therefore needs to be
implemented as an optimization process 
\cite{bienenstock}, and not as a maximization 
process, which would lead to a potentially hazardous 
runaway growth of synaptic strengths.

There are four quadrants in the $2\times2$
Hebbian learning matrix, corresponding to
active / inactive pre- and post-synaptic neurons,
out of which we use the following three
optimization rules:
 
(a) The sum over active afferent links 
leading to active dHan neurons
is optimized to a large but finite value 
$r_v^{act}$,
$$
\sum_{j} v_{ij}\, y_j \bigg|_{x_i\,{\rm active}}
 \ \to\ r_v^{act}~.
$$

(b) The sum over inactive afferent links leading
to active dHan neurons is optimized to
a small value $s_v^{orth}$,
$$
\sum_{j} v_{ij}\,(1-y_j) \bigg|_{x_i\,{\rm active}}
 \ \to\ s_v^{orth}~.
$$

(c) The sum over active afferent links leading to
inactive dHan neurons is optimized to
a small value $r_v^{ina}$,
$$
\sum_{j} v_{ij}\, y_j \bigg|_{x_i\,{\rm inactive}}
 \ \to\ r_v^{ina}~.
$$

The $r_v^{act}$, $r_v^{ina}$ and
$s_v^{orth}$ are the target values for the
respective optimization processes, where the
superscripts stand for `active', `inactive'
and `orthogonal' \cite{gros2008_karl}. These
three optimization rules correspond to
fan-in normalizations of the afferent
synapses. Positive correlations are build up whenever
$r_v^{act}$ dominates in magnitude, and
orthogonalization of the receptive fields
to other stimuli is supported by $s_v^{orth}$.
A small but non-vanishing value for $r_v^{ina}$ 
helps to generate a certain, effective,
fan-out normalization, avoiding the 
uncontrolled downscaling of temporarily
not needed synapses.

\begin{figure*}[tb]
\centerline{
\includegraphics[width=0.75\textwidth]{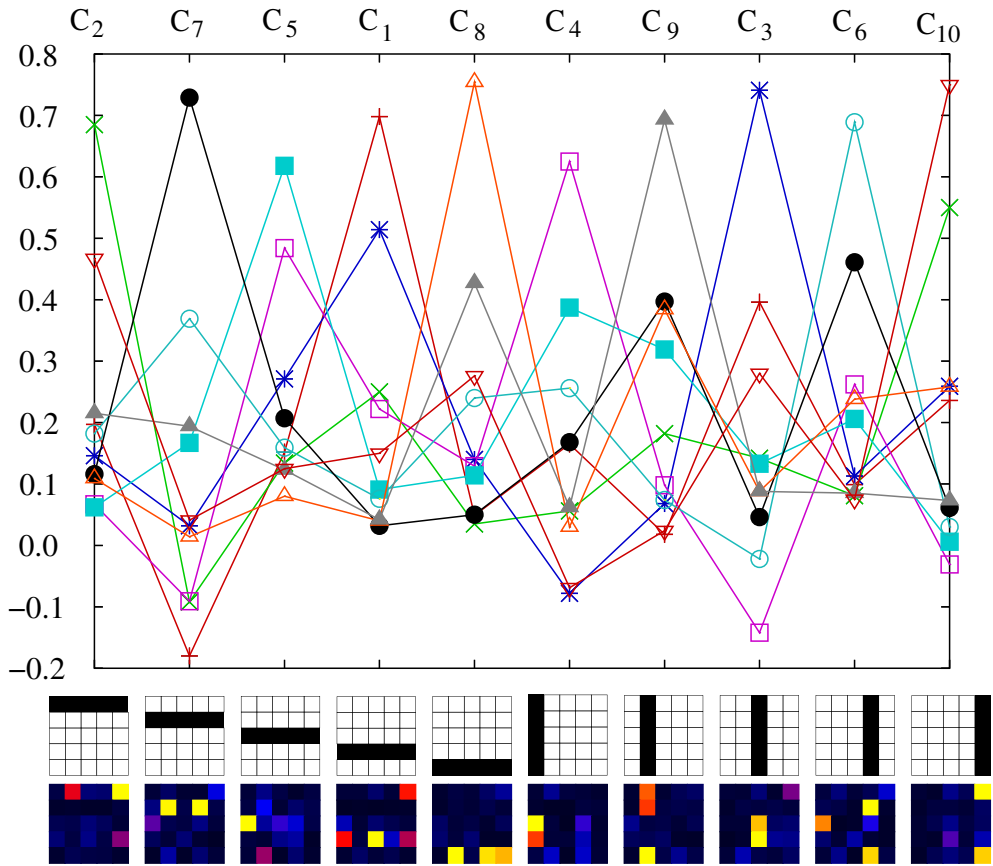}
           }
\caption{
For the $5\times5$ bars problem the response
(see Eq.~(\ref{eq_clique_rec_patt})), of the 10 cliques 
$C_1,..,C_{10}$ in the dHan layer. The clique response
is given with respect to the ten reference patterns, 
{\it viz} the 5 horizontal bars and the 5 vertical bars 
of the $5\times5$ input field shown below the graph.
In the top row the cliques with the maximal response 
to the respective reference patterns
is given. In the bottom row, below each of the
10 black / white reference patterns, the
receptive fields, Eq.~(\ref{eq_clique_recp_fields}),
for the cliques $C_\alpha$ with the maximal susceptibility 
(given in the top row) are shown color-coded, 
with black/blue/red/yellow coding synaptic 
strengths of increasing intensities.
        }
\label{figure_cRP_graph}
\end{figure*}

\subsection{The bars problem}

Knowledge about the environment lies at the basis 
of all cognition, before any meaningful
action can be taken by a cognitive system.
For simple organisms this knowledge is implicitly
encoded in the genes, but in general a cognitive
system needs to extract this information autonomously
from the sensory data input stream, via unsupervised 
online learning. This task 
includes signal separation and features extraction,
the identification of recurrently appearing patterns, 
i.e.\ of objects, in the background of fluctuation
and of combinations of distinct and noisy patterns.
For the case of linear signal superposition this problem
is addressed by the independent component 
analysis \cite{hyvarinen2000}
and blind source separation \cite{choi2005}, which seeks 
to find distinct representations of statistically
independent input patterns.
In order to examine how our system of an input layer coupled
to a dHan layer, as illustrated in Fig.\ \ref{figure_dhan_input},
analyzes the incoming environmental signals,
we have selected the bars problem \cite{barsProblem,butko2007}.

The bars problem constitutes a standard non-linear reference
task for feature extraction via a non-linear independent 
component analysis for an $L\times L$ input layer. The basic patterns
are the $L$ vertical and $L$ horizontal bars and the individual 
input patterns are made up of a non-linear superposition
of the $2L$ basic bars, containing any of them with a 
certain probability $p$, typically $p\approx0.1$, 
as illustrated in Fig.\ \ref{figure_star_pattern}.

Our full system then consist of the dHan layer,
which is continuously active, and an input layer coding
the input patterns consisting of randomly superimposed 
black / white bars. For the dHan network we have
taken a regular 20-site ring, containing a total
of 10 cliques $C_\alpha,\ \alpha=1..10$, 
each clique having $S(C_\alpha)=4$ sites, as illustrated 
in Fig.\ \ref{figure_star_pattern}. The self-sustained
transient-state process is continuously active in
the dHan layer, modulated by the contributions
$\{\Delta r_i\}$ it
receives via the links $v_{ij}$ from the input layer.
For the simulation a few thousands of input patterns
were presented to the system \cite{grosKaczor2008}.

In Fig.\ \ref{figure_cRP_graph} we present 
for the $5\times5$ bars problem the simulation
results for the susceptibility
\begin{equation}
R(\alpha,\beta) \ =\ {1\over S(C_\alpha)}
\sum_{i\in C_\alpha,j} v_{ij}\, y_j^{\beta},
\quad
\begin{array}{rcl}
&&\alpha=1,..,10 \\
&&\beta = 1,..,10
\end{array}
\label{eq_clique_rec_patt}
\end{equation}
of the 10 cliques $C_\alpha$ in the dHan
layer to the 10 basic patterns
$\{y_j^{\beta},j=1,..,25\}$, the 10 individual
horizontal and vertical bars, with
$C_1=(0,1,2,11)$, $C_2=(2,3,4,13)$, and so on.
All cliques have the size $S(C_\alpha)\equiv 4$
and the notation $i\in C_\alpha$ denotes the
set of all sites defining the clique $C_\alpha$.
At the start all $v_{ij}$ are drawn randomly.

The result is quite remarkable. At the beginning
of the simulation the system undergoes an associative
thought process without semantic content. During
the course of the simulation, via the
competitive novelty learning scheme, the individual
attractor relics of the transient state dynamics,
the cliques of the dHan layer, acquire a semantic
connotation, having developed pronounced susceptibilities
to statistically distinct objects in the sensory data 
input stream. This can be seen directly inspecting the 
clique receptive fields
\begin{equation}
F(\alpha,j)\ =\
{1\over S(C_\alpha)} \sum_{i\in C_\alpha} v_{ij},
\qquad \alpha=1,..,10,
\label{eq_clique_recp_fields}
\end{equation}
of the $\alpha=1,...,10$ cliques in the dHan layer with
respect to the $j=1,...,25$ input neurons, which are also
presented in Fig.\ \ref{figure_cRP_graph}. The
clique receptive fields $F(\alpha,j)$
correspond to the averaged receptive fields of 
their constituent neurons. The data presented 
in Fig.\ \ref{figure_cRP_graph} are for the 
$5\times5$ bars problem. We note that simulation 
for larger systems can be performed as 
well, with similar results \cite{grosKaczor2008}.

The learning scheme employed here is based on 
optimization and not on maximization, as stressed in
Sect.\ \ref{subsec_afferent_link_plasticity}.
The clique receptive fields, shown in
Fig.\ \ref{figure_cRP_graph}, are therefore not
of black / white type, but differentiated. 
Synaptic modifications are turned progressively
off when sufficient signal separation has been achieved.
This behavior is consistent with the `learning by error'
paradigm \cite{chialvo1999}, which states that a
cognitive system learns mostly when making 
errors and not when performing well. 

We may take a look at the results presented
in Fig.\ \ref{figure_cRP_graph} from a somewhat
larger perspective. The neural activity of
newborn animals consists of instinct-like reflexes
and homeostatic regulation of bodily functions.
The processing of the sensory signals has not
yet any semantic content and internal neural
activity states do not correspond yet to
environmental features like shapes, colors and
objects. The neural activity can acquire semantic
content, philosophical niceties apart, only through
interaction with the environment. This is a demanding
task, since the optical or acoustical sensory signals
are normally overloaded with a multitude of overlapping
primary objects. The animal therefore needs to
separate these non-linearly superposed signals
for the acquisition of primary knowledge
about the environment and to map the independent signals,
the environmental object to distinct neural activity
patters.

This very basic requirement is performed by the dHan
architecture. The internal transient states have, at
the start of the simulation, no relation to environmental
objects and are therfore void of semantic content.
In the simulation presented here, there are 10 primary 
environmental objects, the 5 horizontal and vertical bars 
of the $5\times5$ bars problem. In the setting used these 
10 objects are independent and statistically uncorrelated. 
During the course of the unsupervised and online learning 
process, the receptive fields of the transiently 
stable neural states, the cliques in the dHan layer, 
acquire distinct susceptibilities not to arbitrary 
superpositions of the primary objects but to the 
individual primary bars themselves.
A sensory signal consisting of the non-linear superposition
of two or more bars will therefore lead, in general,
to the activation of one of the corresponding cliques.
To be concrete, comparing
Fig.\ \ref{figure_cRP_graph}, an input signal
containing both the top-most and the bottom-most
horizontal bar would activate either the clique
$C_2$ or the clique $C_8$. These two cliques
will enter the competition for the next winning
coalition whenever the input is not too weak
and when it overlapps with a sensitive period. The 
present state together with its dynamical
attention field \cite{gros2005} will then determine 
the outcome of this competitions and one of the two objects
present in this input signal is then re\-cognized.

\section{CONCLUSIONS AND DISCUSSION}

The vast majority of neural nets considered to date
for either research purposes, or for applications, are
generalized stimulus-response 
networks \cite{haykin1994,dreyfus2005}. One has 
typically an input signal and an output result,
as, e.g., in speech recognition. In most settings
the network is reset to a predefined default state
after a given task is completed, and before the next
input signal is provided. This approach is
highly successful, in many instances, but it is clearly
not the way the brain works on higher levels. It
is therefore important to examine a range of
paradigmal formulations for the non-trivial 
eigendynamics of cognitive systems, evaluating
their characteristics and computational capabilities.

As an example for a concept situated somewhere
in between a pure stimulus response net and
systems with a fully developed eigendynamics,
we have discussed in Sect.\ \ref{subsec_reservoir_computing} 
the notion of reservoir computing. For reservoir networks
the dynamics is, in general, still induced by the
input signal and decays slowly in the absence of any
input. Any given stimulus encounters however an 
already active reservoir net, with the current reservoir
activity caused by the preceding stimuli. The response
of the network therefore depends on the full history of
input signals and time prediction tasks constitute consequently
the standard applications scenaria for reservoir computing.

A somewhat traditional view, often presumed implicitly,
is that the eigendynamics of the brain results from
the recurrent interlinking of specialized individual cognitive
modules. This viewpoint would imply, that attempts
to model the autonomous brain dynamics can be considered 
only after a thorough understanding of the
individual constituent modules has been achieved.
Here we have examined an alternative route, considering
it to be important to examine the mutual benefits and
computational capabilities of a range of theory proposals 
for the overall organization of the eigendynamics. 

In Sect.\ \ref{subsec_saddle_point_networks} we
have examined a first proposal for the organization
of the eigendynamics in terms of saddle point networks.
In this framework the internal neural dynamics is
guided by heteroclines in a process denoted winnerless
competition. This neural architecture aims to model
reproducible cognitive behavior and a single robust attractor
in terms of a heteroclinic channel constitutes the
eigendynamics in the absence of sensory inputs.

In Sect.\ \ref{subsec_attractor_relic_networks} we
have examined the viewpoint that a non-trivial
associative thought process constitutes the 
autonomous dynamics in the absence of sensory input.
For any finite (and isolated) network these 
thought processes turn eventually into limiting cycles 
of transient states. In this architecture there is
however not a unique limiting cycle, but many
possible and overlapping thought processes, every
one having its respective basin of attractions. The
transient state dynamics required for this approach
is obtained by coupling an attractor network to
slow variables, with the neural time evolution slowing
down near the such obtained attractor relics. This
is a quite general procedure and a wide range of 
concrete implementations are feasible for this concept.

The coupling of neural nets having a non-trivial
eigendynamics to the sensory input is clearly
a central issue, which we have discussed 
in depth in Sect.\  \ref{sect_influence_stimuli},
for the case of networks with transient state
dynamics based on attractor ruins, emphasizing 
two functional principles in this context:

(a) -- The internal transient state dynamics is based
intrinsically on the notion of competitive neural dynamics.
It is therefore consistent to assume that the sensory
input contributes to this neural competition, modulating
the already ongoing internal neural competition. The
sensory input would therefore have a modulating
and not a forcing influence. The sensory signals would
in particular not deactivate a currently stable winning
coalition, influencing however the transition from
one transiently stable state to the subsequent winning
coalition.

(b) -- The eigendynamics of the cognitive system and of the
sensory signals resulting from environmental activities are,
a priori, unrelated dynamically. Correlations between these 
two dynamically independent processes should therefore be 
built up only when a modulation of the internal neural 
activity through the sensory signal has actually occurred. 
This modulation of the eigendynamics by the input data stream
should then generate an internal reinforcement signal, which
corresponds to a novelty signal, as the deviation of the 
internal thought process by the input is equivalent, from 
the perspective of the cognitive system, to something 
unexpected happening.

We have shown, that these two principles can be implemented
in a straightforward manner, resulting in what one could
call an `emergent cognitive capability'. The system performs,
under the influence of the above two general operating
guidelines, autonomously a non-linear independent component
analysis. Statistically independent object in the sensory data 
input stream are mapped during the life time of the cognitive
system to the attractor relics of the transient state
network. The internal associative thought process 
acquires thus semantic content, with the time series of
transient states, the attractor ruins, now corresponding
to objects in the environment.

We believe that these results are encouraging and that
the field of cognitive computation with auto\-no\-mously
active neural nets is an emerging field of growing
importance. It will be important to study alternative
guiding principles for the neural eigendynamics, for the
coupling of the internal autonomous dynamics
to sensory signals and for the decision making
process leading to motor output. Architectures built up 
of interconnected modules of autonomously active neural nets 
may in the end open a pathway towards the development 
of evolving cognitive systems.



\addtolength{\textheight}{-0cm}   

\end{document}